\documentclass[a4paper,12pt]{article}
\usepackage[ansinew]{inputenc}
\usepackage[T1]{fontenc}
\usepackage[english]{babel}
\usepackage{latexsym}
\usepackage{amsmath}
\usepackage{amssymb}
\usepackage[dvips]{graphicx}
\usepackage{epsfig}
\usepackage{subfigure}
\newcommand{\mysection}[1]    
	{                   
	\section{#1}
	 
	}

\newcommand{\lsim}{\buildrel < \over {_\sim}}

\newcommand{\mc}{\mathcal}

\newcommand{\bs}{\boldsymbol}

\newcommand{\eps}{\varepsilon}
\newcommand{\ol}{\overline}
\newcommand{\nl}{\newline}

\newcommand{\beq}{\begin{equation}}
\newcommand{\eeq}{\end{equation}}
\newcommand{\beqa}{\begin{eqnarray}}
\newcommand{\eeqa}{\end{eqnarray}}
\newcommand{\beqan}{\begin{eqnarray*}}
\newcommand{\eeqan}{\end{eqnarray*}}
\newcommand{\nn}{\nonumber}

\newcommand{\bc}{\begin{center}}
\newcommand{\ec}{\end{center}}
\newcommand{\Erf}{\textrm{Erf}}
\newcommand{\Rea}{\textrm{Re}}
\newcommand{\Ima}{\textrm{Im}}

\title{Thermal leptogenesis in a 5D split fermion scenario with bulk neutrinos}
\author{Jukka Maalampi\\ {\small  \textit{Department of Physics, 
University of Jyväskylä, Finland}} \\ {\small and} {\small \textit{Helsinki Institute of Physics, Helsinki, Finland}}\\ {\small and} \\  Iiro Vilja {\small and} Heidi Virtanen\\
 {\small  \textit{Department of Physics and Astronomy, 
University of Turku, Finland}}}
\date{\today}

\begin{document}

\maketitle

We study the thermal leptogenesis in a hybrid model, which combines the so called split fermion model and the bulk neutrino model defined in five dimensional spacetime. This model predicts the existence of a heavy neutrino pair nearly degenerate in mass, whose decays might generate a CP violation large enough for creating the baryon asymmetry of the universe through leptogenesis. We investigate numerically the constraints this  sets on the parameters of the model such as the size of the compactified fifth dimension.

\section{Introduction}
\setcounter{equation}{0}
\setcounter{footnote}{0}

The origin of the baryon-antibaryon asymmetry observed in our Universe is one of the most 
intriguing open questions of the modern cosmology. It is also a question of particle physics as it is
one of the most compelling pieces of evidence of the 
incompleteness of the Standard Model (SM). Had the early universe been 
matter-antimatter symmetric at the temperatures above the electroweak 
phase transition temperature $\mc O$(100 GeV), one would expect the ratio of 
the present number densities of matter over photons to be ${n_B}/{n_\gamma}\simeq 10^{-18}$ (see eg \cite{turner}).
This contradicts the observational result  
\beqa \label{wmapresult}
\frac{n_B}{n_\gamma}=(6.1\pm 0.3)\times 10^{-10}
\eeqa
obtained by the Wilkinson Microwave Anisotropy Probe (WMAP)
\cite{wmap}. It has turned out that in order to explain this huge discrepancy 
one has to consider particle physics models that produce a larger asymmetry between matter and 
antimatter than what is possible to achieve within the SM.

There are three general conditions \cite{sakharov}, called the Sakharov conditions,  that must be fulfilled for a baryon asymmetry to be 
created in the early universe: 
C and CP violation, baryon number (B) violation, and an exit from thermal 
equilibrium.  Several particle physics models have been presented where the Sakharov conditions can be fulfilled without conflicts with other constraints, 
among these are various supersymmetric extensions of the SM (\cite{susyandsakharov} and references therein). The SM is not among these viable models as the baryogenesis would require the Higgs boson to 
be much lighter than the experiments indicate \cite{Kajantie:1995kf}.

One of the most 
appealing scenarios for the creation of the matter-antimatter asymmetry is the baryogenesis via leptogenesis 
(see eg \cite{Fukugita:1986hr},\cite{luty}). In this scheme  one  extends the electroweak sector of the SM with interactions that
violate lepton and baryon number conservation. 
A net lepton number is generated perturbatively e.g.  via triangle diagrams involving lepton number violating couplings, and this lepton number is then converted to a net baryon number by sphalerons through the so 
called Kuzmin-Rubakov-Shaposhnikov (KRS) mechanism \cite{krs}. Several models realizing this 
scenario have been proposed. Most of the contemporary leptogenesis scenarios, 
however, rely on a different mechanism, the so called thermal leptogenesis where a net lepton number is generated via heavy neutrino decays. In this mechanism, proposed in \cite{Fukugita:1986hr}, heavy neutrinos with a mass of the order of the Grand Unification scale  (GUT scale)  undergo CP violating decays producing an lepton-antilepton asymmetry among the decay products. Heavy neutrinos serve also another purpose in these models as they offer, via the see-saw mechanism \cite{mohapatra}, an explanation for the lightness of the  known SM neutrinos.

Another class of models is provided by the so called low energy extra dimension brane models inspired by superstring theories. In these brane-world scenarios our universe is supposed to be a 4 dimensional hypersurface, called the brane, living in a larger dimensional space-time, called the bulk.
It is supposed that the SM particles, including the ordinary left-handed neutrinos, reside on the brane, while sterile particles such as right-handed neutrinos are allowed to propagate also in the bulk \cite{arkanihamedrussell,dienesdudasgherghetta}. The extra dimension theories can offer a new solution to the hierarchy problem by bringing the fundamental 
scale of gravity ($M_{0}$)  many orders of magnitude below
the effective gravity scale, Planck scale $M_{Pl}$. In addition, they might also explain the mass hierarchies within the SM fermion families \cite{arkanihamed,thickbrane}. 

The flavour can be brought into extra dimension models for example by introducing a separate bulk neutrino for each SM neutrino 
\cite{arkanihamedrussell,dienesdudasgherghetta}. This scheme is not restrictive as far as neutrino mixing patterns are concerned allowing for a 
diversity of effective mixing matrices among ordinary neutrinos. Its shortcoming is, however, the great number of undetermined parameters it brings along, which makes its predicting power quite limited.   
Another possibility is to introduce just one type of bulk neutrino with flavour-universal couplings to the SM neutrinos \cite{dienessarcevic}, which means that the coupling between the brane and bulk neutrinos is the same irrespective of the SM flavour of the brane neutrino. It turns out, however, that this kind of scheme would lead to effective mixing matrices that are too rigid for reproducing the  mass and mixing patterns of neutrinos observed in neutrino oscillation experiments. 

An extra dimension model that circumvents  these problems was presented Dienes and Hossenfelder in \cite{dienes}, where the bulk neutrino scheme is combined with the so called split-fermion scenario \cite{thickbrane}. In the split-fermion scenario the SM fermions are each centered in the brane around a different locations and their mixings are due to the overlapping of their corresponding wave functions. The split-fermion scenario at such suffers serious fine tuning problems as the couplings between particles are exponentially sensitive to relative particle distances in the brane. In order the model to reproduce the observed features of neutrino mixing the relative locations of neutrinos on the brane are strictly constrained \cite{Barenboim:2001wy}. In the model proposed in \cite{dienes} such fine tuning problems are avoided. The model is a hybrid model where the split-fermion picture is extended by including bulk neutrinos. It allows the effective neutrino mixing angles to be completely decoupled from the sizes of the wavefunction overlaps on the brane. 

\indent In the present paper we will revive the hybrid model of \cite{dienes} and study it from the point of view of leptogenesis. We will work with a simplified version of the model considering one extra dimension and just one neutrino flavour as flavour does not play any essential role in leptogenesis. Our aim is to investigate whether leptogenesis can be realized in the 
framework of the hybrid model, in particular whether the parameter values required by the leptogenesis scenario are in accordance with the general setup of the model. Finding the allowed size of the extra dimension is key and indicates whether one dimension is sufficient from the beginning or not. This also reveals whether the model can address the hierarchy problem. The plan of the paper is as follows. In Section 2 we will introduce the hybrid model in the form we shall use it. In Section we will consider neutrino phenomenology the model leads to. 
The realization of leptogenesis in the model is presented in Section 4. Section 5 gives a summary of the results and our conclusions.

\section{The Hybrid Model}
\setcounter{equation}{0}
\setcounter{footnote}{0}

In this Section we describe the basic structure of the hybrid model following the original work \cite{dienes}.  The general framework consists of $n_f$
neutrino flavour eigenstates $\Psi_\alpha =(\nu_\alpha,\nu_{\alpha R})^T$ 
($\alpha = 1,\dots ,n_f$) bound to live on the brane and a four-component fermion $\Psi=(\psi_+,\ol\psi_-)^T$ that can propagate in the bulk. Apart from the four-dimensional spacetime there is one extra spatial dimension compactified with a radius $R$. 

It is assumed that each of the active brane neutrinos $\Psi_\alpha$ has a coupling with the bulk neutrino $\Psi$ through a Yukawa term $g\ol \Psi_\alpha H P_R(\Psi+\Psi ^c)$, where $H$ is a Higgs field. At this point we differ from the original model of \cite{dienes} by introducing a complex phase.  We assume that there is a phase difference between the couplings of the the left-chiral $\psi_+$ and the right-chiral $\ol\psi_-$ components of the bulk neutrino $\Psi$. The phase is necessary for the leptogenesis as it allows for CP violation needed for
the creation of a net lepton number in the decays of heavy neutrinos. 

Explicitly, the action on which we will base our analysis is 
written in terms of two-component spinors as follows:
\beqa \label{branebulkaction}
\mc S_c&=&\int d^4xdy\sum_{\alpha=1}^{n_f}
\Big\{M_*\nu^\dagger_\alpha(x,y)\left[\psi_+^c(x,y)
+e^{i\delta_\alpha}\ol\psi_-(x,y)\right]\\
&&+\nu_\alpha ^\dagger(x,y)g\, h(x,y)
\left[\psi_+^c(x,y)+e^{i\delta_\alpha}\ol\psi_-(x,y)\right]\Big\}
+\textrm{h.c.}\nn.
\eeqa
Since we are dealing with a fat brane that is shifted away from orbifold fixed points, the coupling $\nu^\dagger_\alpha (x,y)(M_*e^{i\delta_\alpha}+gh(x,y)e^{i\delta_\alpha})\psi_-(x,y)$ is allowed \cite{dienesdudasgherghetta}. Otherwise, if the brane was located at an orbifold fixed point, then the orbifold boundary conditions would forbid the brane neutrinos $\nu_\alpha$ from coupling to the odd bulk modes of $\psi_-(x,y)$. The universal coupling scale is $M_*=g \langle H\rangle $ and the vacuum expectation value of the 5D Higgs
is written as $\langle H\rangle =v/\sqrt{2\pi R}$,
where 
$g$ is the dimensionful Yukawa coupling in the five-dimensional spacetime and $v$ is the vacuum 
expectation value of the Higgs field in the ordinary four-dimensional spacetime. Hence the universal brane-bulk coupling strength has the  expression $M_*=vg/\sqrt{2\pi R}$. 

For the gamma matrices $\Gamma^A=(\Gamma^{\mu},\Gamma^4)$ ($\mu=0,...,3$)
we use the chiral representation
\beqan
\Gamma^\mu&=&\begin{pmatrix}0&\sigma^\mu\\
\ol \sigma^\mu&0 \end{pmatrix},\qquad \mu=0,...,3\\
\Gamma^4&=&\begin{pmatrix}-i\mathbb I_{2\textrm x2}&0\\
0&i\mathbb I_{2\textrm x2}\end{pmatrix}.
\eeqan

The  kinetic terms of neutrinos in the five-dimensional action are given by
\beqa \label{kineticbulk}
\mc S_\nu&=&\int d^4x dy\sum_{\alpha=1}^{n_f}\nu_\alpha^\dagger(x,y) i\ol 
\sigma^\mu \partial_\mu \nu_\alpha (x,y) ,\\
\mc S_b&=&\int d^4xdy\ol \Psi i \Gamma^A\partial_A \Psi . \nn
\eeqa
The nonzero contribution to the kinetic term of the brane neutrino contains only the 4-dimensional derivatives as the $y$ derivative renders the integrand odd with respect to $y$ and thus the $y$ integral vanishes at its end points. We assume that the extra spatial dimension undergoes an orbifold compactification. By making 
use of the orbifold relations $\psi_+(-y)=\psi_+(y)$ and 
$\ol\psi_-(-y)=-\ol\psi_-(y)$ we can write the Kaluza-Klein (KK) expansions in the following form: 
\beqa
\psi_+(x,y)&=&\frac{1}{\sqrt{2\pi R}}\psi^{(0)}_+(x)
+\frac{1}{\sqrt{\pi R}}\sum_{n>0}\psi^{(n)}_+(x)\cos\frac{ny}{R} , \nn\\
\ol\psi_-(x,y)&=&
\frac{1}{\sqrt{\pi R}}\sum_{n>0}\ol\psi^{(n)}_-(x)\sin\frac{ny}{R}.
\eeqa
\indent For the brane neutrinos located in the fat brane we use the 
Gaussian wave functions
\beqa
\nu_\alpha(x,y)=\frac{1}{\sqrt{\sigma}}\exp\left(-\frac{\pi}{2}
\frac{(y-y_\alpha)^2}{\sigma^2}\right)\nu_\alpha(x) ,
\eeqa
where $\nu_\alpha(x)$ is a four-dimensional spinor. For simplicity we assume that the wave functions of all flavours in the brane have the same width of $\sigma\ll R$. We also 
follow the assumption that the Higgs field profile in the extra dimension is constant 
 \cite{thickbrane} and plug in the zero mode from the Kaluza-Klein expansion:
\beqa
h(x,y)=\frac{1}{\sqrt{2 \pi R}}h(x).
\eeqa
This choice ensures the canonical normalization of the kinetic term of the 
4D Higgs field $h(x)$.

\section{Neutrinos in the hybrid model}
\setcounter{equation}{0}
\setcounter{footnote}{0}

Let us study the neutrino sector of the hybrid model in more detail. Following the original analysis of \cite{dienes} we determine the mass 
spectrum of neutrinos and  the corresponding mass eigenstates.

The mass matrix in four spacetime dimensions is obtained by integrating the actions $S_c$, $S_\nu$ and $S_b$, given in Eqs. (\ref{branebulkaction}) and (\ref{kineticbulk}),
over the extra dimension $y$ leading to
\beqa \label{action}
\mc S_\nu &=&\int d^4x\sum_{\alpha=1}^{n_f}\nu^\dagger_\alpha i\ol \sigma^\mu \partial_\mu \nu_\alpha,\nn\\
\mc S_b&=&\int d^4 x\Big\{\psi^{(0)\dagger}_+ i \ol\sigma^\mu \partial_\mu \psi^{(0)}_++\sum_{n>0}\bigg[\psi^{(n)\dagger}_+ i\ol \sigma^\mu\partial_\mu\psi^{(n)}_++\ol \psi^{(n)\dagger}_- i\sigma^\mu \partial_\mu \ol \psi^{(n)}_-\bigg]\nn\\
&&+\sum_{n>0}\frac{n}{R}\bigg[\psi^{(n)\dagger}_+ \ol \psi^{(n)}_-+\ol \psi^{(n)\dagger}_-\psi^{(n)}_+\bigg]\Big\},\nn\\\mc S_c&=&\int d^4x \sum_{\alpha=1}^{n_f}\Big\{\nu^\dagger_\alpha(x)\bigg[ m\psi^{(0)c}_+(x)+\sum_{n>0}\big ( m^\alpha_{n,+}\psi_+^{(n)c}(x)+m^\alpha_{n,-}\ol\psi^{(n)}_-(x)\big)\bigg]\\
&&+\nu^\dagger_\alpha(x)\bigg[ \frac{hm}{v}\psi^{(0)c}_+(x)+\sum_{n>0}\big( \frac{h(x)m^\alpha_{n,+}}{v}\psi_+^{(n)c}(x)+\frac{h(x)m^\alpha_{n,-}}{v}\ol\psi^{(n)}_-(x)\big)\bigg]\Big\}+\textrm{h.c.},\nn
\eeqa
where $n_f$ is the number of flavours residing on the brane. For the volume-suppressed couplings between the fields on the brane and in the bulk  we have used the following notations:
\beqa \label{couplings}
m&\equiv &M_*\sqrt{\frac{\sigma}{\pi R}}=\frac{gv}{\sqrt{2 \pi R}}\sqrt{\frac{\sigma}{\pi R}},\nn \\
m^ \alpha _{n,+} &\equiv &\sqrt{2} m\cos \left(\frac{ny_\alpha}{R}\right)\textrm{exp}\left [-\frac{n^2\sigma^2}{2\pi R^2}\right ], \nn \\
m^ \alpha _{n,-} &\equiv &\sqrt{2} m e^{i\delta _\alpha}\sin\left(\frac{ny_\alpha}{R}\right)\textrm{exp}\left [-\frac{n^2\sigma^2}{2\pi R^2}\right ].
\eeqa
In what follows we will assume that the brane-bulk coupling is weak and set $mR\ll$1. 

\indent The mass terms appearing  in the action (\ref{action}) are collected together as to
\beqa
\mc S_{\textrm{mass}}&=&\int d^4x\Bigg\{\sum_{\alpha=1}^{n_f}\nu_\alpha^\dagger(x)\Big [m\psi^{(0)c}_0+\sum_{n>0}\bigg (m^\alpha_{n,+}\psi^{(n)c}_+(x)+m^\alpha_{n,-}\psi^{(n)}_-(x)\bigg)\Big]+\textrm{h.c.}\nn\\
&&+\sum_{n>0}\frac{n}{R}\big[\psi^{(n)\dagger}_+\psi^{(n)}_-+\psi^{(n)\dagger}_-\psi^{(n)}_+\big]\Bigg\}.
\eeqa
This can be presented in matrix form as follows: 
\beqa
\mc S_{\textrm{mass}}=\int d^4x\frac{1}{2}(\mc N_\textrm{L}^\dagger \mc M\mc N_\textrm{L}^{c}+\mc N_{\textrm{L}}^{c\dagger}\mc M^*\mc N_\text{L}),
\eeqa
where the mass matrix $\mc M$ is given by
\beqa \label{massmatrix}
\mc M&=&\begin{pmatrix} \mc M_\textrm{L} & \mc M_\textrm{D}\\
\mc M^\textrm{T}_\textrm{D} & \mc M_\textrm{R} \end{pmatrix}\\
\mc M_{\textrm{D}(n_f+1)\times \infty }&=&\begin{pmatrix}m^1_{1,+} & m^1_{1,-}&\dots & m^1_{n,+} & m^1_{n,-}&\dots\\
\vdots & \vdots & \ddots & \vdots & \vdots & \dots\\
m^{n_f}_{1,+} & m^{n_f}_{1,-} & \dots & m^{n_f}_{n,+} & m^{n_f}_{n,-} & \dots\\
0 & 0 & 0 & 0 & 0 & 0 \end{pmatrix},\nn\\
\mc M_{\textrm{L}(n_f+1)\times(n_f+1)}&=& \left(\begin{array}{cc}0 & m\\
m^T & 0\end{array}\right)\nn,\\
\mc M_{\textrm{R}(\infty \times\infty)}&=&\left(\begin{array}{ccccc}0 & \frac{1}{R} & 0 & 0 & \dots\\
\frac{1}{R} & 0 & 0 & 0 & \dots\\
0 & 0 & 0 & \frac{2}{R} & \dots\\
0 & 0 & \frac{2}{R} & 0 & \dots\\
\vdots & \vdots & \vdots & \vdots & \ddots\end{array}\right).\nn
\eeqa

The left- and right-handed fields are arranged into the vectors $\mc N_\textrm{L}$ and $\mc N^c_\textrm{L}$ as follows:
\beqa 
\mc N_\textrm{L}^c&=&(\nu_\alpha^c,\psi^{(0)c}_+,\psi^{(1)c}_+,\psi^{(1)}_-,...,\psi^{(n)c}_+,\psi^{(n)}_-,...)^\textrm{T},\nn\\
\mc N_\textrm{L}&=&(\nu_\alpha,\psi^{(0)}_+,\psi^{(1)}_+,\psi^{(1)c}_-,...,\psi^{(n)}_+,\psi^{(n)c}_-,...)^\textrm{T}.\nn
\eeqa

The matrices $\mc M$ and $\mc M^*$ can be transformed to a block-diagonal form by the transformation \cite{dienes}
\beqa\label{transformation}
T=\begin{pmatrix} I & \kappa \\
-\kappa^\textrm{T} & I \end{pmatrix}
\eeqa
where we have denoted
\beqa
\kappa=\mc M_\textrm{D}\mc M^{-1}_\textrm{R}.
\eeqa
The transformation takes the mass matrix $\mc M$ into the form
\beqa
\widetilde{\mc M}=T^\textrm{T}\mc M T\approx\begin{pmatrix} \widetilde{\mc M}_\textrm{L} & 0 \\
0 & \widetilde{\mc M}_\textrm{R} \end{pmatrix},
\eeqa
where
\beqa \label{blockmass}
\widetilde{\mc M}_\textrm{L}&=&\mc M_\textrm{L}-\kappa \mc M^\textrm{T}_\textrm{D}=\begin{pmatrix} -\sum_n\big (m^\alpha_{n,-}m^\beta_{n,+}+m^\alpha_{n,+}m^\beta_{n,-}\big)\frac{R}{n} & m\\
m & 0 \end{pmatrix}\\
&\equiv & \begin{pmatrix} m_{\alpha\beta} & m\\
m & 0 \end{pmatrix}\nn.
\eeqa


The sum over $n$ in the left upper block $m_{\alpha\beta}$ of the matrix $\widetilde{\mc M}_L$ can be approximated by an integral over $k=(\sigma/R)n$ because the sum is rendered finite as the Gaussian width of the brane neutrinos $\sigma$ acts as a regulator. This results in
\beqa\label{mab}
m_{\alpha\beta}&=&-\frac{M_*^2\sigma}{2}\Big\{e^{i\delta_\beta}\bigg[\Erf\big(\frac{\sqrt{\pi}}{2\sigma}(y_\alpha+y_\beta)\big)-\Erf\big(\frac{\sqrt{\pi}}{2\sigma}(y_\alpha-y_\beta)\big)\bigg]\\
&&+e^{i\delta_\alpha}\bigg[\Erf\big(\frac{\sqrt{\pi}}{2\sigma}(y_\alpha+y_\beta)\big)+\Erf\big(\frac{\sqrt{\pi}}{2\sigma}(y_\alpha-y_\beta)\big)\bigg]\Big\},\nn
\eeqa
where the error function emerges in the integrations over the trigonometric functions appearing in the quatities $m^ \alpha _{n,\pm}$ \cite{dienes}. Upon block-diagonalizing $\mc M^*$ the upper block becomes just the complex conjugate of $\widetilde{\mc M}_{\textrm{L}}$. 

The transformation (\ref{transformation}) renders the field vectors $\mc N_\textrm{L}$ and $\mc N_\textrm{L}^c$ to the form 
\beqa
\widetilde{\mc N_{\textrm{L}}^c}&=&(\widetilde{\nu^c_\alpha},\widetilde{\psi^{(0)c}_+},\widetilde{\psi^{(1)c}_+},\widetilde{\psi^{(1)}_-},...,\widetilde{\psi^{(n)c}_+},\widetilde{\psi^{(n)}_-},...)^\textrm{T}=T^{\textrm{T}}\mc N_{\textrm{L}}^c\\
\widetilde{\mc N_{\textrm{L}}}&=&(\widetilde{\nu_\alpha},\widetilde{\psi^{(0)}_+},\widetilde{\psi^{(1)}_+},\widetilde{\psi^{(1)c}_-},...,\widetilde{\psi^{(n)}_+},\widetilde{\psi^{(n)c}_-},...)^\textrm{T}=T^\dagger \mc N_{\textrm{L}},\nn 
\eeqa
where
\beqa
\widetilde{\nu_\alpha^c}&=&\nu_\alpha^c+\sum_n(m^\alpha_{n,-}\psi^{(n)c}_++m^\alpha_{n,+}\psi^{(n)}_-)\frac{R}{n},\\
\widetilde{\psi^{(0)c}_+}&=&\psi^{(0)c}_+,\nn\\
\widetilde{\psi^{(n)c}_+}&=&\psi^{(n)c}_+-\sum_{\alpha=1}^{n_f}m^\alpha_{n,-}\frac{R}{n}\nu_\alpha^c,\nn\\
\widetilde{\psi^{(n)}_-}&=&\psi^{(n)}_--\sum_{\alpha=1}^{n_f}m^\alpha_{n,+}\frac{R}{n}\nu_\alpha^c\nn
\eeqa
and  
\beqa
\widetilde{\nu_\alpha}&=&\nu_\alpha-\sum_n(m^{\alpha*}_{n,-}\psi^{(n)}_++m^\alpha_{n,+}\psi^{(n)c}_-)\frac{R}{n},\\
\widetilde{\psi^{(0)}_+}&=&\psi^{(0)}_+,\nn\\
\widetilde{\psi^{(n)}_+}&=&\psi^{(n)}_++\sum_{\alpha=1}^{n_f}m^{\alpha*}_{n,-}\frac{R}{n}\nu_\alpha,\nn\\
\widetilde{\psi^{(n)c}_-}&=&\psi^{(n)c}_-+\sum_{\alpha=1}^{n_f}m^\alpha_{n,+}\frac{R}{n}\nu_\alpha.\nn
\eeqa


\indent We proceed  by determining the eigenvalues and eigenvectors of the neutrino mass matrix in the simplified case where we take into account the 
electron neutrino  only and assume that the muon and tau neutrinos are decoupled, in other words, the mixing angles between the electron neutrino and other active neutrinos in the brane are assumed to be zero. In order to make the scheme to work in the case of one extra dimension,  two left-handed brane neutrinos are needed: in addition to the SM electron neutrino $\nu_e$, there must exist a left-handed sterile neutrino. Introducing the additional brane neutrino ensures we get a spectrum where one mass eigenstate is light while the remaining two are heavy. In relation to sterile neutrinos, a recent neutrino experiment suggests there are no light sterile neutrinos \cite{sterilenuexperiment}. In our case the extra brane neutrino is related to a heavy mass eigenstate. The requirement of a plausible mass spectrum in terms of leptogenesis compels us to set the number of neutrino flavours to two, that is $n_f=2$.  
 
\indent With the above assumption, the mass matrix (\ref{blockmass}) becomes a 3$\times$3 matrix ($n_f=2$) of the form
\beqa \label{upperblock}
\widetilde{\mc M}_L=\left( \begin{array}{ccc} m_{11} & m_{12} & m \\
m_{12} & m_{22} & m\\
m & m & 0 \end{array} \right),
\eeqa
where $m$ is defined in Eq.(\ref{couplings}) and $m_{\alpha\beta}$ in Eq.(\ref{mab}).
The eigenvalues of $\widetilde{\mc M}_\textrm{L}$ are
\beqa
\lambda_1&\simeq & \frac{1}{2}(m_{11}+m_{22})-m_{12},\\
\lambda_2&\simeq & -\sqrt{2} m+\frac{1}{2}m_{12}+\frac{1}{4}(m_{11}+m_{22}),\nn\\
\lambda_3&\simeq & \sqrt{2} m+\frac{1}{2}m_{12}+\frac{1}{4}(m_{11}+m_{22})\nn
\eeqa
and those of $\widetilde{\mc M}_\textrm{L}^*$ are just $\lambda_1^*$, $\lambda_2^*$ and $\lambda_3^*$. Due to the complex nature of $m_{\alpha\beta}$, these eigenvalues are generally complex. The physical masses are $m_1=|\lambda_1|$, $m_2=|\lambda_2|$ and $m_3=|\lambda_3|$. One has $m_1<<m_2\approx m_3$.
The corresponding mass eigenstates are given by the following superpositions of the interaction eigenstates: 
\beqa
\chi_1&=&\frac{1}{\sqrt{2}}e^{i\theta_1/2}\Big (-\frac{1}{\sqrt{2}}\frac{m_{11}-m_{22}}{|m_{11}-m_{22}|}\widetilde{\nu^c_1}+\frac{1}{\sqrt{2}}\frac{m_{11}-m_{22}}
{|m_{11}-m_{22}|}\widetilde{\nu^c_2}\Big)\nn\\
&&+\frac{1}{\sqrt{2}}e^{-i\theta_1/2}\Big(-\frac{1}{\sqrt{2}}\frac{m^*_{11}-m_{22}^*}{|m_{11}-m_{22}|}\tilde\nu_1
+\frac{1}{\sqrt{2}}\frac{m^*_{11}-m^*_{22}}{|m_{11}-m_{22}|}\tilde \nu_2\Big)\\
\chi_2&=&\frac{1}{\sqrt{2}}e^{i\theta_2/2}\Big(-\frac{1}{2}\widetilde{\nu^c_1}-\frac{1}{2}\widetilde{\nu^c_2}+\frac{1}{\sqrt{2}}\psi^{(0)c}_+\Big)+\frac{1}{\sqrt{2}}e^{-i\theta_2/2}\Big(-\frac{1}{2}\tilde\nu_1-\frac{1}{2}\tilde\nu_2+\frac{1}{\sqrt{2}}\psi^{(0)}_+\Big)\nn\\
\chi_3&=&\frac{1}{\sqrt{2}}e^{i\theta_3/2}\Big(-\frac{1}{2}\widetilde{\nu^c_1}-\frac{1}{2}\widetilde{\nu^c_2}+\frac{1}{\sqrt{2}}\psi^{(0)c}_+\Big)+\frac{1}{\sqrt{2}}e^{-i\theta_3/2}\Big(-\frac{1}{2}\tilde\nu_1-\frac{1}{2}\tilde\nu_2+\frac{1}{\sqrt{2}}\psi^{(0)}_+\Big),\nn
\eeqa
where $\theta_i={\rm arg}(\lambda_i)$. The complex factors  in $\chi_{1,2,3}$ will give rise to the desired CP asymmetry in the decays of the heavy states $\chi_{2,3}$.


\section{CP Violation and Leptogenesis}
\setcounter{equation}{0}
\setcounter{footnote}{0}

Let us move to study leptogenesis in the model described above. As was mentioned, an attractive scenario for creating the baryon asymmetry 
consists of generating a lepton number asymmetry through lepton number violating decays of heavy neutrinos, followed by the generation of baryon 
asymmetry from this lepton asymmetry via anomalous {\it B+L} conserving  effects during the electroweak phase transition.  

The CP-violation in heavy neutrino decays arises in leading order through the interference of the tree level amplitude and the lowest order 
vertex corrections \cite{Fukugita:1986hr,luty,sarkar,covi,buchjaplu}. It has been shown, however, that in some cases the interference of the tree-level amplitude with the diagram  where one heavy state is transformed to the other via light lepton and Higgs loop (called as the mixing amplitude), will give a major contribution to the CP violation \cite{liusegre,sarkarwf,coviwf,pilaftsiscp}. This may happen 
if the states that mix are a pair of nearly degenerate heavy neutrinos. As we have seen, in the model we are interested in there is an almost degenarete neutrino pair $\chi_{2}$, $\chi_{3}$. We will therefore concentrate in what follows on the CP-violation arising from the interference of the tree level digram and the mixing diagram. 

\begin{figure}[ht]
\centering
\subfigure[]{
\includegraphics[scale=1]{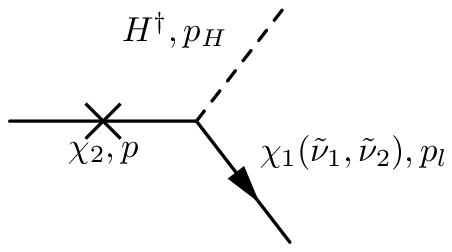}
\label{fig:subfig1}
}
\quad
\subfigure[]{
\includegraphics[scale=0.9]{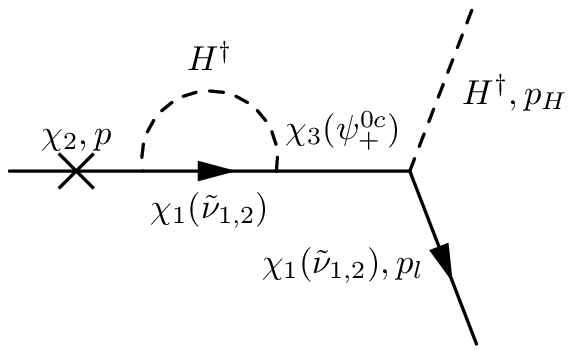}
\label{fig:subfig2}
}
\quad
\subfigure[]{
\includegraphics[scale=0.9]{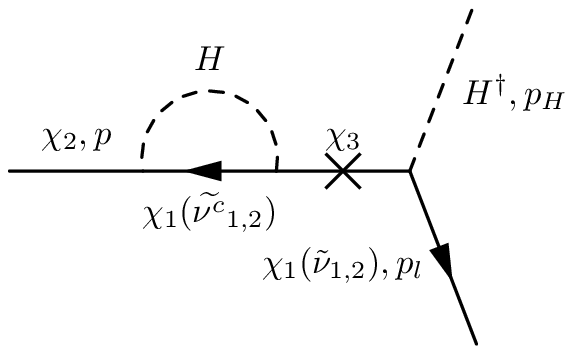}
\label{fig:subfig3}
}
\caption{The relevant Feynman diagrams for the process $\chi_2\rightarrow \chi_{1\textrm{L}} H^\dagger$. The tree level diagram due to the decay of $\chi_2$ to a neutrino and Higgs is in Fig \ref{fig:subfig1}. Fig \ref{fig:subfig2} and \ref{fig:subfig3} depict the mixing diagrams due to the the decay of $\chi_2$ to a light neutrino and Higgs. The mass insertion occurs prior and after the loop, respectively.}
\label{fig:leptondecays}
\end{figure}

\begin{figure}[ht]
\centering
\subfigure[]{
\includegraphics[scale=1]{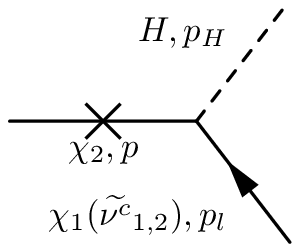}
\label{fig:antisubfig1}
}
\quad
\subfigure[]{
\includegraphics[scale=0.9]{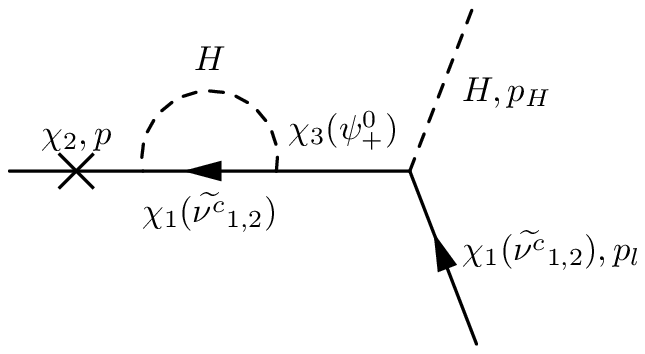}
\label{fig:antisubfig2}
}
\quad
\subfigure[]{
\includegraphics[scale=0.9]{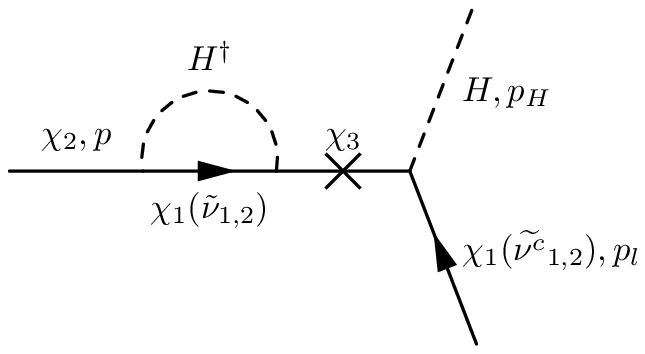}
\label{fig:antisubfig3}
}
\caption{The relevant Feynman diagrams for the process $\chi_2\rightarrow \chi_{1\textrm{R}} H$. The tree level diagram due to the decay of $\chi_2$ to a light antineutrino and Higgs is in Fig \ref{fig:antisubfig1}. Fig \ref{fig:antisubfig2} and \ref{fig:antisubfig3} depict the mixing diagrams due to the the decay of $\chi_2$ to a light antineutrino and Higgs. The mass insertion occurs prior and after the loop, respectively.}
\label{fig:antileptondecays}
\end{figure}

Let us study the decays of $\chi_2$. The Feynman diagrams relevant from the point of view of CP violation are those presented in Figures \ref{fig:leptondecays} 
and \ref{fig:antileptondecays}. Fig \ref{fig:subfig1} depicts the tree level decay of $\chi_2$ into a light neutrino and Higgs boson, and  Figs \ref{fig:subfig2} and \ref{fig:subfig3} present  one loop diagrams where the process proceeds  through a transition of $\chi_2$ into an intermediate $\chi_3$. The corresponding diagrams for antineutrino production in the decays of $\chi_2$ are presented in Figs. \ref{fig:antisubfig1}, \ref{fig:antisubfig2} and \ref{fig:antisubfig3}. The diagrams for the decays of $\chi_3$  are obtained  from those of $\chi_2$ decays, presented in the figures, by interchanging
$\chi_2$ and $\chi_3$.

As the heavy neutrinos in the model we are looking at are nearly degenerate, their mass difference being $|m_2-m_3|\sim M_*^2\sigma\ll m$, we can expect the interference 
between the tree level diagrams of Figs
\ref{fig:subfig1} and \ref{fig:antisubfig1} and the one-loop mixing diagrams of Figs \ref{fig:subfig2}, \ref{fig:subfig3}, \ref{fig:antisubfig2} 
and \ref{fig:antisubfig3} to give the leading contribution to the CP violation in neutrino decays.  

The CP violation arises from the difference between the decay widths of the lepton and antilepton production channels. In terms of the amplitudes, the relevant quantity is
\beqan
|\mc M_0+\mc M_{1}|^2-|\ol{\mc M}_0+\ol{\mc M}_{1}|^2\simeq 2 \Rea(\mc M^*_0\mc M_{1})-2\Rea (\ol{\mc M}^*_0\ol{\mc M}_{1})
\eeqan
where $\mc M_0$ and $\mc M_{1}$ are the tree level amplitude and mixing amplitude, respectively and $\ol{\mc M}_0$ and $\ol{\mc M}_{1}$ 
denote the corresponding antiparticle amplitudes.
The CP-asymmetry parameter  that takes into account the decay of both heavy mass eigenstates can be defined as 
\cite{pilaftsiscp}
\beqa \label{totalepsilon}
\eps=\frac{\Gamma(\chi_2\rightarrow H^\dagger \chi_{1\textrm{L}})-\ol\Gamma (\chi_2\rightarrow H\chi_{1\textrm{R}})+
\Gamma(\chi_3\rightarrow H^\dagger \chi_{1\textrm{L}})-\ol\Gamma (\chi_3\rightarrow H\chi_{1\textrm{R}})}
{\Gamma(\chi_2\rightarrow H^\dagger \chi_{1\textrm{L}})+\ol\Gamma (\chi_2\rightarrow H\chi_{1\textrm{R}})+
\Gamma(\chi_3\rightarrow H^\dagger \chi_{1\textrm{L}})+\ol\Gamma (\chi_3\rightarrow H\chi_{1\textrm{R}})}.
\eeqa
(Another definition can be found in \cite{sarkar} where asymmetries are calculated separately for $\chi_2$ and $\chi_3$ and then added together. )

The phenomenological constraints of the value of the CP-violation parameter $\epsilon$ are obtained by relating it to the observed baryon asymmetry of the universe.  The lepton number generated via heavy lepton decays is related to the CP violation parameter $\eps$ through (see eg. \cite{Hubble rate} and \cite{Fukugita:1986hr})
\beqa
Y_L\simeq \kappa\frac{\eps}{g_*}.
\eeqa
Here the parameter $\kappa$ is a factor that describe the dilution of the lepton asymmetry due to various lepton number conserving and violating processes taking place in the primordial plasma. Its value is estimated to be $\kappa=10^{-2}-10^{-1}$ \cite{Hubble rate, kappafactor, leptogenesislecture}.
The lepton number $Y_L$ created is partially transformed to  baryon number $Y_B$ due to anomalous electroweak processes. The lepton number and the net baryon number are related  through \cite{blnumbers}
\beqa
Y_B=\frac{n_B}{s}=\frac{c_s}{c_s-1}Y_L=\frac{c_s}{c_s-1}\frac{n_L}{s},
\eeqa
where $s$ is entropy, $n_B$ and  $n_L$ are the baryon and lepton number density, respectively. In our case the factor $c_s$ is 
\beqa
c_s&=&\frac{8n_f+4}{22n_f+13}= \frac{20}{57},
\eeqa
yielding $Y_B=-{20}Y_L/{37}$. Taking the observational value for the baryon asymmetry, as given in Eq. (\ref{wmapresult}), and the relation $s=7.04 n_\gamma$ between the entropy $s$ and the photon number density $n_\gamma$, we will arrive at the condition  
\beqa
6.1\times 10^{-10}=\frac{n_B}{n_\gamma}=7.04\kappa\frac{c_s}{c_s-1}\frac{\eps}{g_*}.
\eeqa
Given the estimated values for the parameter $\kappa\sim 0.01 \-- 0.1$, we the obtain the following order-of-magnitude estimation for the allowed values of the parameter $\epsilon$: 
\beqa \label{leptogenesiscondition}
-\eps \simeq  10^{-7}\-- 10^{-6}.
\eeqa
Since we work in a regime where the SM particles have acquired masses via the electroweak symmetry breaking, the sphaleron transition is not as efficient as it would be in the symmetric phase. Thus (\ref{leptogenesiscondition}) corresponds to the highest amount of CP-violation possible to produce in the model.
 
When calculating the amplitudes, we work in the on-shell renormalization scheme and so the real (dispersive) part of the mixing/self-energy loop vanishes when the propagator mass coincides with the renormalized mass $m_{2,3}$. The couplings between $\widetilde\nu_{1,2}^{c}$ and $\tilde \nu_{1,2}$, which are allowed by the structure of the effective theory (\ref{blockmass}), can be neglaected in the leading order as the Yukawa couplings between the brane neutrinos are very small,  of the order $m_{\alpha\beta}/v\ll m/v$. 

The mixing amplitude of Figs \ref{fig:subfig2} and \ref{fig:subfig3} is given by
\beqa
i\mc M_1^{\chi_2}&=&\frac{1}{2 \sqrt{2}}\frac{m}{v}\frac{m_{11}^*-m_{22}^*}{|m_{11}-m_{22}|}\big (e^{i(\theta_3-\theta _2)/2}-e^{i(\theta_3-\theta _1)/2}\big )\times\\
&&\ol u_lP_{\textrm{R}}u_2\frac{A^*_{32}m_2^2+m_2m_3A_{32}-iA_{33}A^*_{32}m_2^2}{m_2^2-m_3^2-|A_{33}|^2m_2^2-2im_2^2
\Rea A_{33}},\nn
\eeqa
where
\beqa
A_{32}&=&\frac{1}{256\pi}\frac{m^2}{v^2}\Big (\frac{m_{11}-m_{22}}{|m_{11}-m_{22}|}\Big )^2\big (e^{i(\theta_1-\theta_2/2-\theta_3/2)}
+e^{i(\theta_1-\theta_3)/2}\big ),\\
A_{33}&=&\frac{1}{256\pi}\frac{m^2}{v^2}\Big (\frac{m_{11}-m_{22}}{|m_{11}-m_{22}|}\Big )^2\big(e^{i(\theta_1-\theta_3)}+e^{i(\theta_1/2+\theta_2/2-\theta_3)}\big).\nn
\eeqa
The antilepton decay mixing amplitude (Figs \ref{fig:antisubfig2} and \ref{fig:antisubfig3}) is given by
\beqa
i\ol{\mc M}_1^{\chi_2}&=&\frac{1}{2\sqrt{2}}\frac{m}{v}\frac{m_{11}-m_{22}}{|m_{11}-m_{22}|}\big (e^{i(\theta _2-\theta_3)/2}-e^{i(\theta _1-\theta_3)/2}\big )\times\\
&&\ol u_lP_{\textrm{L}}u_2\frac{A_{23}^*m_2^2+m_2m_3A_{23}-iA_{33}A_{23}^*m_2^2}{m_2^2-m_3^2-|A_{33}|^2m_2^2-2im_2^2\Rea A_{33}},\nn
\eeqa
where $A_{23}=A_{33}$. The corresponding mixing amplitudes due to the decay of $\chi_3$ are
\beqa
i\mc M_1^{\chi_3}&=&\frac{1}{2\sqrt{2}}\frac{m}{v}\frac{m_{11}^*-m_{22}^*}{|m_{11}-m_{22}|}\big(1-e^{i(\theta_2-\theta_1)/2}\big)\times\nn\\
&&\ol u_lP_\textrm{R}u_3\frac{A^*_{23}m_3^2+m_2m_3A_{23}-im_3^2A_{22}A_{23}^*}{m_3^2-m_2^2-m_3^2|A_{22}|^2-2im_3^2\Rea A_{22}},\\
i\ol{\mc M}_1^{\chi_3}&=&\frac{1}{2\sqrt 2}\frac{m}{v}\frac{m_{11}-m_{22}}{|m_{11}-m_{22}|}\big(1-e^{i(\theta_1-\theta_2)/2}\big)\times\nn\\
&&\ol u_lP_{\textrm{L}}u_3\frac{m_3^2A^*_{32}+m_2m_3A_{32}-im_3^2A_{22}A^*_{32}}{m_3^2-m_2^2-|A_{22}|^2m_3^2-2im_3^2\Rea A_{22}},\nn
\eeqa
where $A_{22}=A_{32}$.

We find  the following lengthy expression for the CP-violation parameter $\eps$ defined in (\ref{totalepsilon}):
\beqa \label{cpviolationparameter}
\eps &=&\frac{1}{2}(m_2+m_3)^{-1}\Bigg\{\big[(m_2^2-m_3^2-|A_{33}|^2m_2^2)^2+4m_2^4(\Rea A_{33})^2\big]^{-1}m_2\times\nn\\
&&\Bigg [\cos\frac{\theta_3-\theta_2}{2}\Big[(m_2^2-m_3^2-|A_{33}|^2m_2^2)(m_2m_3(\Ima A_{33}-\Ima A_{22})\nn\\
&&+m_2^2(\Rea A_{33}
\Rea A_{22}+\Ima A_{33}\Ima A_{22}-\Rea A_{33}\Rea A_{33}-\Ima A_{33}\Ima A_{33}+\Ima A_{22}-\Ima A_{33}))\nn\\
&&+2m_2^2\Rea A_{33}
(m_2^2(-\Ima A_{33}\Rea A_{22}
+\Rea A_{33}\Ima A_{22}-\Rea A_{22}+\Rea A_{33})\nn\\
&&+m_2m_3(\Rea A_{33}-\Rea A_{22}))\Big ]\nn\\
&&+\sin\frac{\theta_2-\theta_3}{2}\Big[(m_2^2-m_3^2-|A_{33}|^2m_2^2)(m_2m_3(\Rea A_{22}+\Rea A_{33})\nn\\
&&+m_2^2(\Ima A_{33}\Rea A_{22}-\Rea A_{33}\Ima A_{22}+\Rea A_{22}+\Rea A_{33}))\nn\\
&&+2m_2^2\Rea A_{33}(-m_2m_3(\Ima A_{22}+\Ima A_{33})+m_2^2(\Rea A_{33}\Rea A_{22}\nn\\
&&+\Ima A_{33}\Ima A_{22}
+\Rea A_{33}\Rea A_{33}+\Ima A_{33}\Ima A_{33}+\Ima A_{22}+\Ima A_{33}))\Big ]\Bigg ]\\
&&+\big[(m_3^2-m_2^2-|A_{22}|^2m_3^2)^2+4m_3^4(\Rea A_{22})^2\big]^{-1}m_3\times\nn\\
&&\Bigg[\cos\frac{\theta_2-\theta_3}{2}\Big[(m_3^2-m_2^2-|A_{22}|^2m_3^2)(m_2m_3(\Ima A_{22}-\Ima A_{33})\nn\\
&&+m_3^2(\Rea A_{22}\Rea A_{33}+\Ima A_{22}\Ima A_{33}-\Rea A_{22}\Rea A_{22}-\Ima A_{22}\Ima A_{22}+\Ima A_{33}-\Ima A_{22}))\nn\\
&&+2m_3^2\Rea A_{22}(m_2m_3(\Rea A_{22}-\Rea A_{33})+m_3^2(-\Rea A_{33}\Ima A_{22}\nn\\
&&+\Rea A_{22}\Ima A_{33}-\Rea A_{33}+\Rea A_{22}))\Big ]\nn\\
&&+\sin\frac{\theta_3-\theta_2}{2}\Big[(m_3^2-m_2^2-|A_{22}|^2m_3^2)(m_2m_3(\Rea A_{22}+\Rea A_{33})\nn\\
&&+m_3^2(\Ima A_{22}\Rea A_{33}-\Rea A_{22}\Ima A_{33}+\Rea A_{33}+\Rea A_{22}))\nn\\
&&+2m_3^2\Rea A_{22}(-m_2m_3(\Ima A_{33}+\Ima A_{22})+m_3^2(\Rea A_{22}\Rea A_{33}\nn\\
&&+\Ima A_{22}\Ima A_{33}+\Rea A_{22}\Rea A_{22}+\Ima A_{22}\Ima A_{22}+\Ima A_{33}+\Ima A_{22}))\Big]\Bigg]\Bigg\}\nn
\eeqa
We have treated oth the mass difference $|m_2-m_3|$ and the Yukawa coupling squared $m^2/v^2$ as perturbation variables as they are roughly of the same order of magnitude and both small in comparison with the masses of the decaying neutrinos:
\beqa \label{perturbation}
\frac{m²}{v²}m_{2,3}\sim |m_2-m_3| \ll m_2, m_3.
\eeqa
The corrections of first order in these parameters are taken into account in (\ref{cpviolationparameter}), higher order corrections are negligible.

A few comments concerning our result are in order. The expression which we have obtained  for the CP violation parameter $\epsilon$ is more complicated than, for example, the expressions obtained in minimal models \cite{sarkarwf,coviwf, pilaftsiscp} based on SO(10) GUT. This is mainly due to the fact that in the model we are considering, based on the assumption of an extra spatial dimension, the CP-violation is generated in an energy scale where all fermions have achieved their mass as a result of the electroweak symmetry breaking. As a consequence, both brane and bulk neutrinos contribute to the mass matrix. Hence, our light mass eigenstate involves both two brane neutrino states, which results in a more involved combination of the Yukawa couplings $(A_{ij})$ in the mixing loop. For example, in \cite{pilaftsiscp} the factors fulfill $A_{ij}=A^*_{ji}$ and thus terms $\mc O(A_{ij}^2)$ cancel, whereas in our case $A_{ij}$'s are not symmetric ($A_{ij}\ne A^*_{ji}$) and the cancellation does not occur.

In order to have a viable mechanism for the creation of the baryon asymmetry, all three Sakharov conditions have to be fulfilled. The third condition requires that the expansion rate of the universe, given by the Hubble parameter $H(T)$, must be greater than the tree-level 
decay rate of any $L$-violating process. This condition will in our case set a constraint on the 5D Higgs vacuum expectation value $v$.

The dominant $L$ violating processes in the present model are the heavy neutrino decays considered above. The third Sakharov condition then requires that the decay rates obey the condition (heavy neutrino mass denoted by $m_N$)
\beq
\label{firstcondition} 
\Gamma^{\textrm{tree}}\lsim 2 H(T=m_{N}),
\eeq
which guarantees that heavy neutrinos are out of equilibrium when they decay. The tree-level decay rate is easily calculated to be (see eg \cite{Hubble rate} for decay rates of heavy particles and CP asymmetry producation)
\beq
\Gamma^{\textrm{tree}}=\frac{m_N}{64\pi}\frac{m^2}{v^2}\bigg(1-\cos\frac{\theta_2-\theta_1}{2}\bigg),
\eeq
 and the Hubble rate at the decoupling of the heavy neutrinos is given by \cite{Hubble rate}
\beq
H(T=m_{N})=1.73\sqrt{g_*}\frac{m_N^2}{M_{Pl}},
\eeq
where $g_*$ is the effective number of degrees of freedom at the stage of heavy neutrino decoupling. One can take $g_*\sim 100$. The condition (\ref{firstcondition}) becomes
\beqa
\frac{m^2}{v^2}&<&3.46\times 64 \pi\sqrt{g_*}\frac{m_N}{M_\textrm{Pl}}\bigg(1-\cos\frac{\theta_1-\theta_2}{2}\bigg)^{-1}.
\eeqa

\begin{figure}[ht]
\centering
\includegraphics[scale=1]{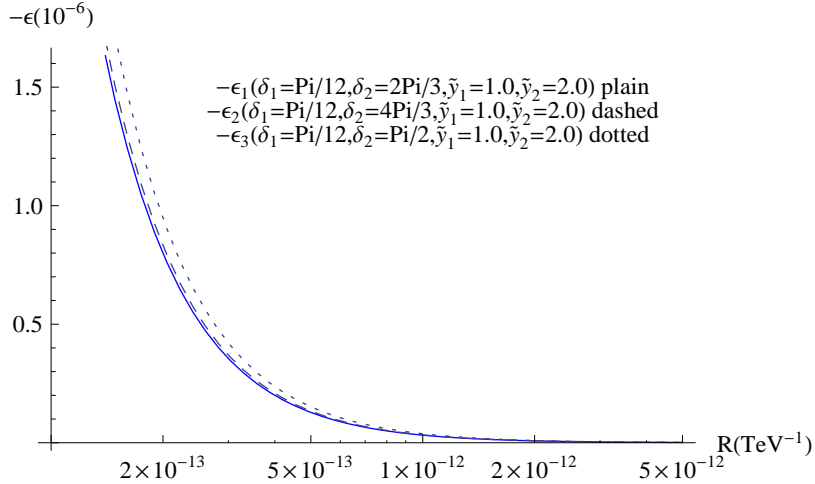}
\caption{The CP violation parameter $\eps$ as a function of the size of the extra dimension $R$, with $-\eps_1(\tilde y_1=1.0$,$\tilde y_2=2.0$,$\delta_1=\pi/12$,$\delta_2=2\pi/3)$,  $-\eps_2(\tilde y_1=1.0$,$\tilde y_2=2.0$,$\delta_1=\pi/12$,$\delta_2=4\pi/3)$,  $-\eps_3(\tilde y_1=1.0$,$\tilde y_2=2.0$,$\delta_1=\pi/12$,$\delta_2=\pi/2)$.}
\label{plotone}
\end{figure}

The CP-violating parameter $\eps$ depends on parameters $\tilde y_{1,2}=\sqrt{\pi}/(2\sigma)y_{1,2},\ \delta_{1,2}$, $R$ and the scale of the light neutrino mass $M_*^2\sigma$ which we take to be $\sim$ 1 eV \cite{lightneutrinomass}.  In Fig \ref{plotone} we present the values of  $\eps$ as a function of  the size $R$ of the extra dimension for three sets of constant values of  the parameters $\tilde y_{1,2}$ and $\delta_{1,2}$. The solid curve in Fig \ref{plotone} corresponds to the set $\tilde y_1=1.0$, $\tilde y_2=2.0$, $\delta_1=\pi/12$ and $\delta_2=2\pi/3$, the dashed curve to the set $\tilde y_1=1.0$, $\tilde y_2=2.0$, $\delta_1=\pi/12$ and $\delta_2=4\pi/3$, and the dotted curve to the set $\tilde y_1=1.0$, $\tilde y_2=2.0$, $\delta_1=\pi/12$ and $\delta_2=\pi/2$.
Comparison of the plot with the constraint (\ref{leptogenesiscondition}) implies that  the allowed size range for the extra dimension for the given set of  parameter values is $R\simeq 2.0\times 10^{-13}\textrm{TeV}^{-1}-4.0\times 10^{-12}\textrm{TeV}^{-1}$. Varying the phase angles $\delta_{1,2}$ to different quadrants  shifts the allowed values of $R$, and the plausible range for $R$ is roughly from $10^{-16}\textrm{TeV}^{-1}$ to $10^{-11}\textrm{TeV}^{-1}$. Hence, sufficient CP violation can be produced in this model when the size of the extra dimension is in the range from roughly the Planck scale to five orders of magnitude larger than the Planck length and the masses of the heavy neutrinos are of the order of $\sim$ 1 TeV. 

\begin{figure}[ht]
\centering
\includegraphics[scale=1]{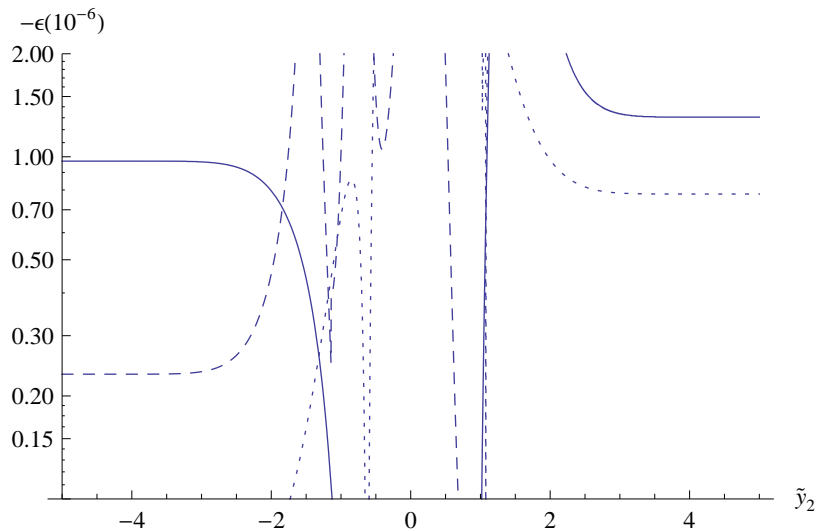}
\caption{The CP violation parameter $\eps$ as a function of the brane location $\tilde y_2$ for $\tilde y_1=1.0$, $\delta_1=\pi/12$, $\delta_2=2\pi/3$ and $R=10^{-13}$ TeV$^{-1}$) (solid curve); $\tilde y_1=1.0$, $\delta_1=\pi/12$, $\delta_2=\pi/3$ and $R=10^{-11}$ TeV$^{-1}$ (dashed curve); $\tilde y_1=1.0$, $\delta_1=\pi/12$, $\delta_2=7\pi/6$ and $R=10^{-11}$ TeV$^{-1}$ (dotted curve).}
\label{plottwo}
\end{figure}

In Fig \ref{plottwo}, the CP-violating parameter $\eps$ is plotted as  as a function of the (normalized) brane location $\tilde y_2$ for three sets of values of the parameters $\delta_{1,2}$, $\tilde y_1$ and $R$. The solid curve corresponds to the set $\delta_1=\pi/12,\ \delta_2=2\pi/3,\ \tilde y_1=1.0,\ R=10^{-13}$ $\textrm{TeV}^{-1}$, the dashed curve to the set $\delta_1=\pi/12,\ \delta_2=\pi/3,\ \tilde y_1=1.0,\ R=10^{-11}$ $\textrm{TeV}^{-1}$, and the dotted curve to the set $\delta_1=\pi/12,\ \delta_2=7\pi/6,\ \tilde y_1=1.0,\ R=10^{-11}$ $\textrm{TeV}^{-1}$. We can see from this plot that the values of the brane location of the neutrino $\nu_2$ that lead to an acceptable vales of $\eps$ depend quite strongly on the values of the phase angles $\delta_i$. 

\begin{figure}[ht]
\centering
\includegraphics[scale=1]{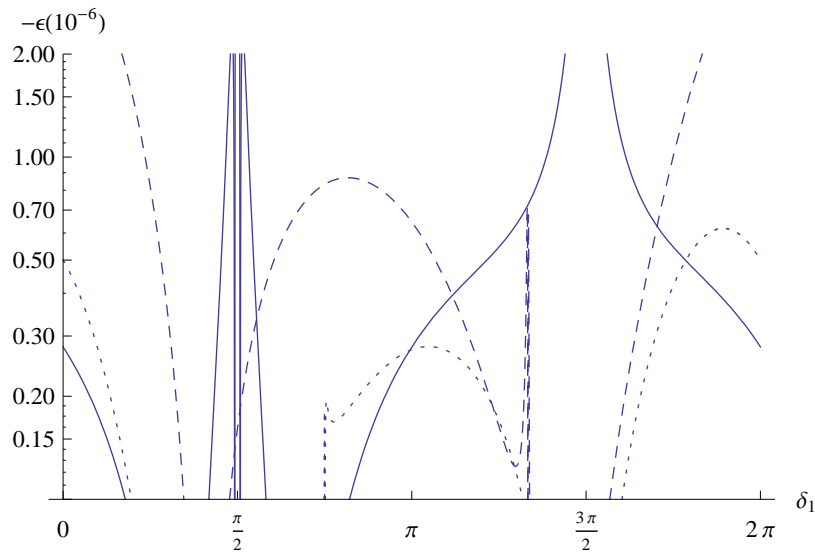}
\caption{The CP violation parameter $\eps$ as a function of the phase angle $\delta_1$, when ($\tilde y_1=1.0$, $\tilde y_2=2.0$, $\delta_2=\pi/2$ and $R=10^{-11}$ TeV$^{-1}$) (solid), ($\tilde y_1=1.0,\ \tilde y_2=2.0,\ \delta_2=4 \pi/3,\ R=10^{-11}$ TeV$^{-1}$) (dashed), ($\tilde y_1=1.0,\ \tilde y_2=2.0,\ \delta_2=3 \pi/4,\ R=10^{-13}$ TeV$^{-1}$) (dotted).}
\label{plotthree}
\end{figure}

Fig \ref{plotthree} presents the CP violation parameter $\epsilon$ as a function of the phase angle $\delta_1$ for three sets of parameters. The solid curve corresponds to the set ($\tilde y_1=1.0,\ \tilde y_2=2.0,\ \delta_2=\pi/2,\ R=10^{-11}$ TeV$^{-1}$), the dashed curve to the set  ($\tilde y_1=1.0,\ \tilde y_2=2.0,\ \delta_2=4\pi/3,\ R=10^{-11}$ TeV$^{-1}$), the dotted curve to the set ($\tilde y_1=1.0,\ \tilde y_2=2.0,\ \delta_2=3 \pi/4,\ R=10^{-13}$ TeV$^{-1}$). 
The allowed values of the brane locations $\tilde y_{1,2}$ and phase angles $\delta_{1,2}$ are quite restricted overall as only highly limited intervals of $\tilde y_{1,2}$ and $\delta_{1,2}$ with any given $R$ lead to $\eps$  of correct order of magnitude and correct sign. However, letting $\delta_{1,2}$ vary leads to a periodic pattern of the values of $\eps$ that are acceptable.

\section{Conclusions}
\setcounter{equation}{0}
\setcounter{footnote}{0}

We have investigated, from the point of view of the leptogenesis, a model with one extra spatial dimension. The model, originally presented in \cite{dienes}, combines the so called bulk neutrino model and the split neutrino model. In this hybrid model different neutrino flavours are assumed to be in separate locations in a thick four-dimensional brane and in bulk there reside sterile neutrinos that couple with these brane neutrinos. Our study shows that in this model a CP violation large enough for leptogenesis to work can be created through decays of heavy neutrinos. We found that the size of the extra dimension should be in the range $10^{-16}$ $\textrm{TeV}^{-1}$ to $10^{-11}$ $\textrm{TeV}^{-1}$ in order to  ensure the correct magnitude of 
CP violation. 

A few concluding remarks are in order. The leading contribution to the CP violation arises from the amplitudes where the tree level diagram interferes with a one-loop self-energy digram where there is a transition between two almost degenerate heavy neutrinos.  The effect of such amplitudes on the CP violation has been earlier studied eg. in \cite{pilaftsiscp}, where it was found that the tree level decay width removes the singularity that occurs when the two heavy neutrinos are degenerate. In our model the coupling structure is different from that of the model of \cite{pilaftsiscp}, and in our case the cancellation of higher order Yukawa terms do not occur but, on the other hand, these terms are perturbatively small.

Extending the analysis of our model to the cases of more than one extra dimensions could also be worthwhile because then the sizes of the extra dimensions could be larger than the one found here. Also, as noted in \cite{dienes}, adding more dimensions would essentially change the neutrino mass spectrum. Namely, the brane-bulk coupling $m$ would be suppressed relative to the brane-brane couplings $m_{\alpha\beta}$, which gives $n_f$ mass eigenstates with masses $\sim M_*^2\sigma$ and one exceedingly light sterile state with  a mass of the order of magnitude of $ \sigma/R^2$. This hierarchy is likely to lead to different constraints on the parameters compared to the ones found in this paper.  

Finally,  we have restricted the analysis to the electron
neutrino only, but we expect the conclusions would be qualitatively similar in the case of other neutrino types.  

\bigbreak

{\bf Acknowledgements} We thank Tuomo Puumalainen for assistance at an early stage of this work. One of us (HV) expresses her gratitude to the Magnus Ehrnrooth Foundation and the Finnish Academy of Science and Letters (the Väisälä Fund) for financial support.

\end{document}